\documentclass[aip,jcp,preprint,superscriptaddress,floatfix,byrevtex]{revtex4-1} 

\usepackage{bm}
\usepackage{amsmath}
\usepackage{amsfonts}
\usepackage{amssymb}
\usepackage{graphicx}
\usepackage{setspace} 

\begin{document}

\title{Solid-state chemistry of glassy antimony oxides}

\author{Chang-Eun \surname{Kim}}
\affiliation{Global E$^3$ Institute and Department of Materials Science and Engineering, Yonsei University, Seoul, Korea}

\author{Jonathan M. \surname{Skelton}}
\affiliation{Centre for Sustainable Chemical Technologies and Department of Chemistry, University of Bath, Bath, UK}

\author{Aron \surname{Walsh}}
\email[Electronic mail: ]{a.walsh@bath.ac.uk}
\affiliation{Centre for Sustainable Chemical Technologies and Department of Chemistry, University of Bath, Bath, UK}
\affiliation{Global E$^3$ Institute and Department of Materials Science and Engineering, Yonsei University, Seoul, Korea}

\author{Aloysius \surname{Soon}}
\email[Electronic mail: ]{aloysius.soon@yonsei.ac.kr}
\affiliation{Global E$^3$ Institute and Department of Materials Science and Engineering, Yonsei University, Seoul, Korea}

\date{\today}

\begin{abstract}
The amorphous phases of antimony oxide at three different oxidation levels have been investigated by {\it ab initio} molecular-dynamics calculations using a simulated `melt-quench' approach. The atomic and electronic structure of the amorphous phases are analyzed and compared to their crystalline counterparts. The amorphous structure of the trioxide ({\it a\,}--Sb$_2$O$_3$) resembles the $\beta$-phase of the crystalline form (valentinite), in agreement with previous observations. In the relaxed athermal structure, however, more senarmontite-like structural features are apparent. The phase diagram of the amorphous phases with respect to oxygen chemical potential has been calculated within the framework of {\it ab initio} thermodynamics. At elevated oxidation levels, the resulting tetraoxide and pentoxide materials show distinct variation in electronic structure compared to the trioxide, with the electronic density of states indicating a narrowing of the electronic gap with increasing oxidation level.
\end{abstract}

\maketitle

\section{Introduction}
Heavy-metallic oxide (HMO) glasses have attracted much attention due to their unique non-linear optical properties.\cite{som2010} Their rapid dielectric-response times, on the order of femtoseconds, are much sought after for use in high-speed optical devices, an application for which homogeneous amorphous materials are ideal candidates. 

The antimony oxides, Sb$_x$O$_y$, have been known as glass formers since the 1930s.\cite{zachariasen1932} There have been experimental studies of glassy antimony oxides using both glass stabilizers\cite{dubois1986,sahar1992,kumar1982} and vitrifying matrices.\cite{terashima1996} Several problems are yet to be solved, however, in particular the issue that the non-linear optical properties of antimony-oxide glasses tend to degrade rapidly as the proportion of the vitrifying matrix increases,\cite{terashima1996} and anionic glass stabilizers (e.g. halides\cite{dubois1986,sahar1992} or phosphates\cite{kumar1982}) tend to result in the formation of microcrystallites and/or to lead to local variation in chemical composition.\cite{nalin2001} In order to overcome the present challenges, it is desirable to investigate high-purity amorphous phases of antimony oxides. In this study, we provide theoretical predictions of the properties of a series of single-component amorphous antimony oxides at different oxidation levels.

In order to understand the local and long-range structure of glassy materials, it is useful to use the crystalline phases with equivalent composition as a reference. Whereas glass materials do not possess a regular atomic geometry, the local atomic arrangements tend to resemble those found in the crystalline counterparts.\cite{terashima1996} In accordance with the stoichiometries of known crystalline antimony oxides, we examine three different compositions of amorphous antimony oxide. We refer to these compositions as {\it a\,}--Sb$_2$O$_x$, where {\it a} denotes an amorphous phase, and $x=3, 4, 5$ indicate the oxidation level.

Crystalline forms of antimony oxide with two different Sb oxidation states are known. Sb(III) and Sb(V) form {\it c\,}--Sb$_2$O$_3$ and {\it c\,}--Sb$_2$O$_5$, respectively, and a mixed phase containing equal proportions of Sb(III) and Sb(V), {\it c\,}--Sb$_2$O$_4$, also exists (the notation {\it c} denotes a crystalline phase).

Senarmontite ($\alpha$--Sb$_2$O$_3$)is the stable phase of {\it c\,}--Sb$_2$O$_3$ under ambient conditions, while valentinite ($\beta$--Sb$_2$O$_3$) is observed at elevated temperature and pressure. In the senarmontite structure, covalently-bonded Sb$_4$O$_6$ units are closely packed in a face-centered cubic (FCC) lattice. The building block of these units is the pyramidal SbO$_3$ motif, in which Sb occupies the center of an elongated tetrahedron with O at three vertex points and the lone-pair electrons of Sb(III) occupying the fourth. The Sb$_4$O$_6$ motif consists of four of these pyramids forming a closed cage-like structure. Valentinite is also composed of SbO$_3$ pyramids, but instead of forming a caged structure, they form a more highly-connected double-chain structure, which transforms to a double-wire structure under pressure. A notable difference between $\alpha$- and $\beta$--Sb$_2$O$_3$ is also evident in the bond-angle distribution function, i.e. a histogram of the three-atom bond angles in the structure. The SbO$_3$ pyramidal unit in senarmontite gives rise to a characteristic O--Sb--O bonding angle of 95.7$^{\circ}$, whereas the arrangement in valentinite leads to three different O--Sb--O angles. 
A third phase, $\gamma$, with regular three-dimensional connectivity, also exists, and it is of interest to see how this structural variation is reflected in the glass network in {\it a\,}--Sb$_2$O$_3$.

With elevated oxygen partial pressure and high temperature, antimony pentoxide (Sb$_2$O$_5$) becomes stable up to about 1250\,K.\cite{orosel2005} The lone-pair electrons are not present in Sb$_2$O$_5$, and instead six-coordinate oxygen atoms form a slightly distorted octahedral arrangement around Sb(V) ions (O--Sb--O bonding angle: 169.2$^{\circ}$). At an intermediate oxidation level, $\alpha$--Sb$_2$O$_4$ and $\beta$--Sb$_2$O$_4$ become stabilized, exhibiting structural features from Sb$_2$O$_5$ (octahedral geometry around Sb(V)) as well as features from Sb$_2$O$_3$ (tetrahedral geometry around Sb(III)). One notable difference between the $\alpha$- and $\beta$- phases is the 'distortion' of the structure around the central Sb(V) ions - whereas the motifs in the $\beta$ phase possess an undistorted octahedral O--Sb--O bond angle of 180$^{\circ}$, the angle in the $\alpha$ phase is closer to 167.5$^{\circ}$. Aside from this distorted bond angle, however, the two phases share a similar atomic structure, and are known to co-exist in the phase diagram up to about 620\,K.\cite{orosel2005}

The antimony oxide series thus encompasses six distinct stoichiometric crystal structures. The interaction between the Sb 5$s$, Sb 5$p$ and O 2$p$ orbitals is known to play an important role in this rich structural chemistry. The diverse range of crystal structures and their physical properties have been characterized experimentally by Orosel and coworkers.\cite{orosel2005,orosel2012} A recent work by Allen {\it et al.}\cite{allen2013} presented a comprehensive theoretical study of the structural, electronic, and optical properties of the known crystalline phases of antimony oxides. However, there are as yet few theoretical studies on the amorphous phases of antimony oxide, despite their unique properties and potential applications.

Quenching single-component, high-purity glassy phases of antimony oxides, and characterising their local structure, still remains a significant experimental challenge. Moreover, the effect of elevated oxidation level on the amorphization of antimony oxides has not been investigated in detail, with the majority of the experimental work having focussed on the lower oxidation state of Sb (i.e. compositions close to {\it a\,}--Sb$_2$O$_3$). It was initially suggested that {\it a\,}--Sb$_2$O$_3$ may have a similar local structure to that of valentinite.\cite{hasegawa1978} However, Terashima {\it et al.} compared the IR and Raman spectra of senarmontite, valentinite and amorphous antimony oxide, and observed vibrational features from both crystalline forms.\cite{terashima1996}

While amorphous antimony oxides at higher levels of oxidation have not been thoroughly examined experimentally, there are several interesting observations which suggest the potential importance of controlling the oxidation level. In one study, an attempt to synthesize glassy antimony oxide led to a mixed-oxidation phase containing both Sb${^{3+}}$ and Sb${^{5+}}$, confirmed with X-ray fluorescence spectroscopy, which is a unique feature of the tetraoxide phase.\cite{masuda1995} Also, Missana {\it et al.} examined the effect of oxygen content on the crystallisation of amorphous antimony oxide, and found that it played a critical role in the kinetics, with a higher oxygen content increasing the thermal energy required to induce crystallisation. This indicates that oxidation level may be an important factor in dictating the (thermal) stability of the glass. 

While precise control of oxidation level during experimental syntheses may be challenging, computational techniques such as {\it ab initio} molecular dynamics (AIMD) based on density-functional theory (DFT) can be a highly-effective tool for studying amorphous materials.\cite{walsh2009,skelton2014} AIMD 'melt and quench' simulations allow atomistic models of glasses with defined compositions to be generated and characterised, which can assist the interpretation of experimental results, and is particularly valuable in cases where spectroscopy has limited utility in probing the detailed atomic structure. In this study, we employ AIMD simulations to generate atomistic models of amorphous antimony oxides with the three oxidation levels observed in {\it c\,}--Sb$_2$O$_x$ {\it viz.} $x$ = 3, 4, and 5. We investigate their fundamental physical properties, including the atomic structure of the glass network, the characteristic local and long-range order, and the electronic structure, and also the thermodynamics of their formation.

This understanding of the structure-property relationships should serve as an important reference point for a detailed understanding of the fundamental properties of the amorphous antimony oxide system.

\section{Methodology}
Periodic electronic-structure calculations were performed within Kohn-Sham DFT\cite{kohn1965} using the Perdew-Burke-Ernzerhof generalized-gradient approximation functional\cite{perdew1996} (PBE-GGA), as implemented in the Vienna {\it Ab initio} Simulation Package (VASP) code.\cite{kresse1996,kresse1996a} 
The kinetic-energy cutoff for the plane-wave basis was set to 500\,eV, and the core electrons were treated using the projector augmented-wave (PAW) method.\cite{blochl1994,kresse1999} The electronic wavefunctions were evaluated at the $\Gamma$ point.

The Nos\'{e}-Hoover thermostat was used to control the temperature,\cite{nose1984} with a fictitious mass parameter corresponding to approximately 40 timesteps (80 fs). The Nos\'{e} mass parameter affects the magnitude of the temperature fluctuations, which in this study were observed to be as large as 7\,\%, measured by the standard deviation of the fluctuations.

Grimme's DFT-D2 scheme\cite{grimme2006} was employed to account for the weak van der Waals forces via a London dispersion correction. Accounting for dispersion forces has been shown to be important especially for the senarmontite structure, in which Sb$_4$O$_6$ clusters are connected by weak interactions. Without a correction for weak forces, the bare PBE-GGA predicts the valentinite to be energetically the most stable phase, contradicting experimental observations,\cite{orosel2012} and the DFT-D2 correction resolves this problem (see Supporting Information). We note that there is a newer DFT-D3 scheme,\cite{grimme2010} which allows variations of the correction parameters depending on the local atomic arrangement of the model. The effect of the structure-dependency of van der Waals correction for an amorphous structure can be an interesting topic for a future study.

Amorphous models consisting of 150, 120 and 140 atoms for S{\it a\,}--Sb$_2$O$_3$, {\it a\,}--Sb$_2$O$_4$ and {\it a\,}--Sb$_2$O$_5$, respectively, were generated by a melt-quench AIMD approach, e.g. as used by Vanderbilt {\it et al.}\cite{vanderbilt2005} In each simulation, an initial configuration generated by seeding atoms onto a cubic grid was prepared. Specifically, the starting atomic configurations of {\it a\,}--Sb$_2$O$_3$, {\it a\,}--Sb$_2$O$_4$, and {\it a\,}--Sb$_2$O$_5$ consist of 60 Sb and 90 O atoms, 40 Sb and 80 O atoms, and 40 Sb and 100 O atoms, respectively. These were modelled in cubic cells of length 13.76, 11.83, and 12.03\,{\AA}, accordingly. The initial structures were randomized for 10\,ps at 3000\,K, and then maintained as a liquid at 1000\,K for a further 20\,ps. This method is advantageous over a simple randomization, since it captures the physical interactions between atoms at elevated temperature. The resulting liquid structures were then quenched to 300\,K over 20 ps, giving a cooling rate of 35\,K\,ps$^{-1}$. A uniform time step of 2\,fs was used, for a total simulation time of 50\,ps, with each trajectory therefore consisting of 25,000 timesteps. 

The MD simulations were carried out at constant volume, with the density set to the average density of the crystalline phases reduced by 5\,\%, as suggested in Ref.\,\onlinecite{vanderbilt2005}. The canonical ({\it NVT}) ensemble, in which the volume of the cell is fixed, avoids some complexities associated with the {\it NPT} ensemble, in particular the slow equilibration of the various barostats commonly used in such simulations. Amorphous materials are typically, although not always, less dense than their crystalline counterparts, and, as noted by Vanderbilt and coworkers\onlinecite{vanderbilt2005}, constraining the density to too high a value can prohibit amorphization. Therefore, we ran the melt-quench AIMD simluations at a lower density than the crystalline phases, and then subsequently volume-relaxed the quenched structures. After the simulated melt-and-quenching process followed by volume relaxation, the calculated density of the amorphous models were 5.59\,g/cm$^3$ for {\it a\,}--Sb$_2$O$_3$, 6.05\,g/cm$^3$ for {\it a\,}--Sb$_2$O$_4$, and 6.01\,g/cm$^3$ for {\it a\,}--Sb$_2$O$_5$. Note that the density of amorphous sample is found to be less denser in experimental study, e.g. 5.05\,g/cm$^3$ for {\it a\,}--Sb$_2$O$_3$ in Johnson {\it et al.}\cite{johnson2003} The internal coordinates and cell parameters of the final quenched structures were relaxed to minimize the residual stress, with convergence criteria of 10$^{-4}$ and 10$^{-5}$\,eV for the total energy and electronic wavefunctions, respectively.

In typical melt-quench experiments, the cooling rate is about 10$^6$\,K s$^{-1}$ for thin-film amorphization, and can be as slow as 100\,K s$^{-1}$ during the formation of bulk-metallic glasses.\cite{shen1999} 
However, a very high cooling rate of 10$^{13}$\,K s$^{-1}$ has been reported to occur during sonochemical synthesis of amorphous materials.\cite{suslick1991} Although there is no literature at present demonstrating the application of this technique to amorphous antimony oxides, the technique has been successfully used to prepare amorphous FeSb$_2$O$_4$ nanoparticles.\cite{nag2011} The timescale of the quenching procedure in the present study is thus comparable to that in at least some experimentally-accessible amorphization techniques.

The radial-distribution function (RDF), $g(r)$, is a commonly-used technique to characterise the bonding structure of amorphous structures:
\begin{equation}
g(r)=\frac{dn}{4\pi r^2 dr} \frac{1}{\rho}\quad;\quad \rho=\frac{N}{V} \quad,
\end{equation}
where $n$ is the number of the ion-ion pairs separated by the distance $r$, $N$ is the number of atoms in the model and $V$ is the volume of the simulation cell. From the point of view of a reference atom, $g(r)$ expresses the probability of finding a neighbour at a distance $r$ relative to that expected from a homogeneous distribution of atoms given the density of the system. The function peaks at preferred interatomic distances, and converges to unity at large $r$ for disordered systems. When the number of ions in the model is small (as is usually the case in AIMD simulations), functions such as this tend to show discrete peaks that would be broadened out by the broader range of variation in local structure in larger systems and/or due to thermal motion at finite temperature. To mimic these effects, the individual distances were convoluted with a Gaussian function of width 0.1\,{\AA} when constructing the histogram.

Another standard structural characterization method for amorphous materials is the bond-angle distribution function, which counts the occurrence of three-atom bond angles and peaks at angles characteristic of the local geometry (e.g. 90$^{\circ}$ for octahedral or square-planar coordination, or ~109$^{\circ}$ for tetrahedral coordination). In the present calculations, we employed a distance cutoff of 2.4\,{\AA}, chosen to be coincident with the first minimum in the RDFs, to define the first coordination shell for identification of bonds. As for the RDFs, each angle was convoluted with a Gaussian function with a width of 1.5$^{\circ}$ before being binned into the histogram. For the calculations of both RDFs and BADs, we had to choose an atomic configuration corresponding to the denoted temperature. We note that the corresponding temperature of the atomic configuration is determined by a linear interpolation of the control temperatures of the MD calculations.

The formation energies, $E_{\rm f}$, of the amorphous oxides were calculated with respect to the chemical potential of oxygen, and compared to those of the crystalline counterparts at different stoichiometries. Bulk antimony was taken as a reference to calculate relative formation energies, $\Delta E_{\rm f}$, according to:
\begin{equation}
\Delta E_{\rm f}(\Delta \mu_{\rm O}) = \frac{\left( E^{\rm tot} - n_{\rm Sb} E^{\rm tot}_{\rm Sb} - n_{\rm O} \Delta \mu_{\rm O} \right)}{n_{\rm Sb}} - E_{\rm f, Sb}    \quad,
\end{equation}
$E^{\rm tot}$ is the total energy of the system, $n_{\rm Sb}$ is the number of antimony atoms, $E^{\rm tot}_{\rm Sb}$ is the total energy per atom of bulk Sb, $n_{\rm O}$ is the number of oxygen ions in the system, $\Delta\mu_{\rm O}$ is the chemical potential of oxygen referenced to the oxygen molecule, and $E_{\rm f, Sb}$ is the formation energy of bulk antimony, which is zero here due to the choice of reference state. The chemical potential of oxygen is calculated according to Richter {\it et al.},\cite{richter2014} with the experimental oxygen-binding energy of 5.22\,eV being used to correct for the error in the PBE value.

Finally, the electronic densities of states (DOS) for the volume-relaxed amorphous structures were calculated by performing single-point calculations with the HSE06 hybrid exchange-correlation functional.\cite{heyd2003,heyd2004,heyd2006,krukau2006,paier2006} For these calculations, a larger $\Gamma$-centred 2$\times$2$\times$2 Monkhorst-Pack {\bf $k$}-point grid was used to sample the Brillouin zone.\cite{monkhorst1976} A linear tetrahedron method with Bl\"{o}chl correction was used for the electronic structure calculations.\cite{blochl1994b} We also performed comparative electronic-structure calculations on the crystalline antimony oxides. The atomic structures of the crystalline antimony oxides are taken from the inorganic crystal structure database (ICSD).\cite{belsky2002} Using these atomic structures, we have conducted self-consistent geometric relaxation calculations using the PBE functional, followed by single-point calculations with the HSE06 functional. Detail computational settings for the {\bf k}-points, and use of van der Waals correction (PBE-D2) are listed in the supporting information.

\section{Results and Discussion}
\begin{figure}[tb!]
\center
\includegraphics[width=0.45\textwidth]{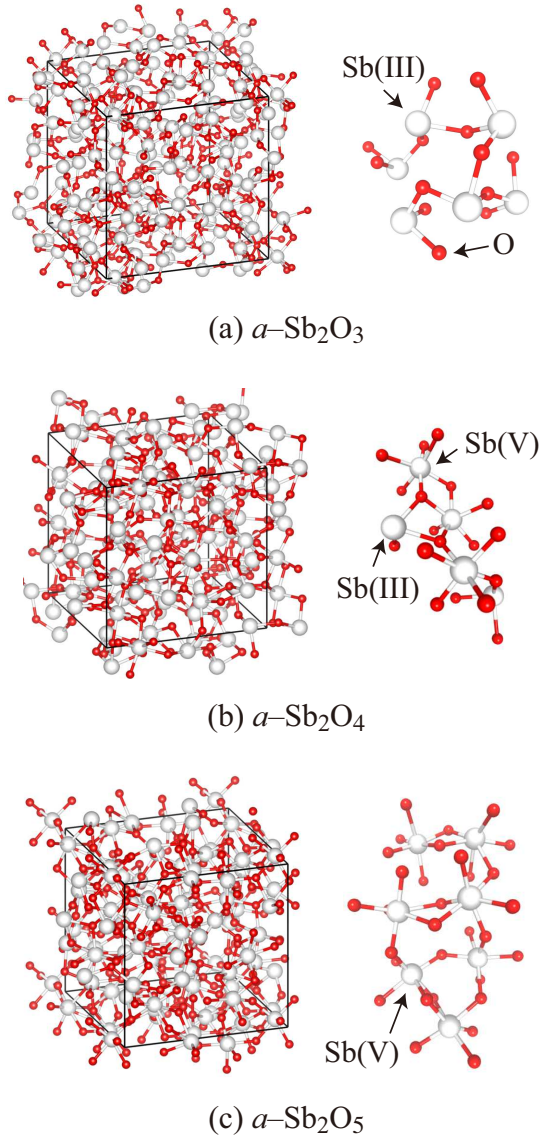}
\caption{(Color online) Atomic structures of the single-component amorphous phases of Sb$_2$O$_3$ (a), {\it a\,}--Sb$_2$O$_4$ (b) and {\it a\,}--Sb$_2$O$_5$ (c). The quenched configuration obtained from the simulations is shown on the left, with representative fragments illustrating the local structure of the glass network to the right.}
\label{figure01}
\end{figure}

The relaxed atomic configurations of the amorphous antimony oxides are shown in Fig.\,\ref{figure01}. All three structures clearly show the absence of long-range order, which is characteristic of amorphous materials, while the glass network can be seen to consist of locally-ordered atomic motifs. 
A common structural feature is that Sb(III) ions display asymmetric four-coordinate geometries comprised of neighboring oxygen atoms and a stereochemically-active lone pair of electrons, while Sb(V) ions adopt slightly distorted octahedral coordination environments.
Following the revised lone pair model,\cite{walsh2011stereochemistry} Allen and co-workers noted that the lone pair in Sb occupies the antibonding 5$s^*$ orbital, and that the oxygen 2$p$ orbital plays a critical role in this electronic configuration, which accounts for the differences in atomic structure around the Sb(III) and Sb(V) ions.\cite{allen2013} 

\begin{figure}[tb!]
\center
\includegraphics[width=0.65\textwidth]{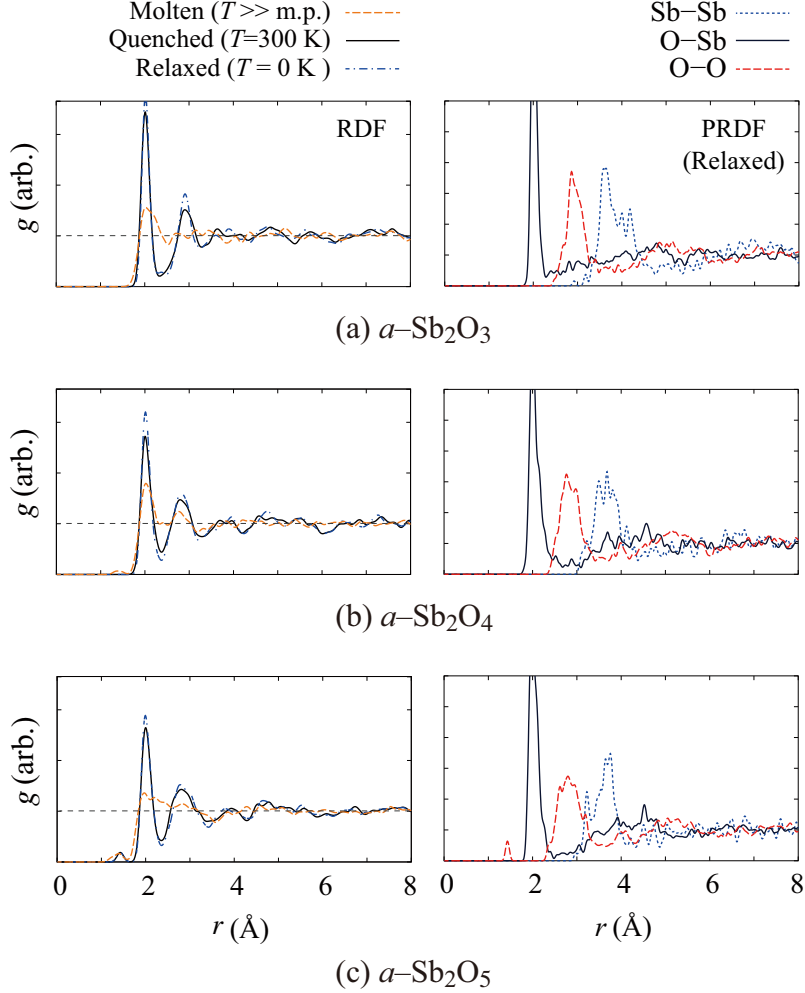}
\caption{(Color online) Calculated radial-distribution functions (RDFs) of amorphous antimony-oxide models with varying oxygen composition, at various temperatures modelled in the calculations. The pairwise partial radial-distribution functions (PRDFs), which describe distance correlations between specific atomic species, are shown in the plots in the right-hand column. The RDFs were calculated from the amorphous structures obtained after the randomization, quenching and relaxation steps, corresponding to the molten, quenched, and athermal relaxed states, respectively. The PRDFs were calculated for the final athermal relaxed structures.}
\label{figure02}
\end{figure}

To obtain more quantitative information about the local structure, we also calculated the total and partial RDFs (PRDFs) of the final configurations obtained at the end of the liquid and quench parts of the MD trajectories, and of the volume-relaxed amorphous models (Fig.\,\ref{figure02}). Another clear indication of the absence of long-range order in the models can be seen from the convergence of the function to unity at $r > 5.0$\,{\AA}. The PRDFs likewise show that there is no long-range order among the various atomic species. The Sb--O RDF peaks around 2.0\,{\AA}, which is in line with the experimental characterization of Johnson {\it et al.},\cite{johnson2003} where simultaneous analysis of neutron and x-ray diffraction (XRD) data gave an Sb-O bond length of 1.99\,{\AA} in amorphous Sb$_2$O$_3$ stabilized by ZnCl$_2$. In the experimental work, the correlation functions of XRD and neutron diffraction surprisingly well matched with the second and the third peaks of the RDF of {\it a\,}--Sb$_2$O$_3$, which indicates not only the nearest Sb--O bonding distances of the model are matching to the experimental characterization, but also the mid- and long-range atomic arrangement of the model are in a reasonable agreement with the experimental data.

\begin{figure}[tb!]
\center
\includegraphics[width=1.00\textwidth]{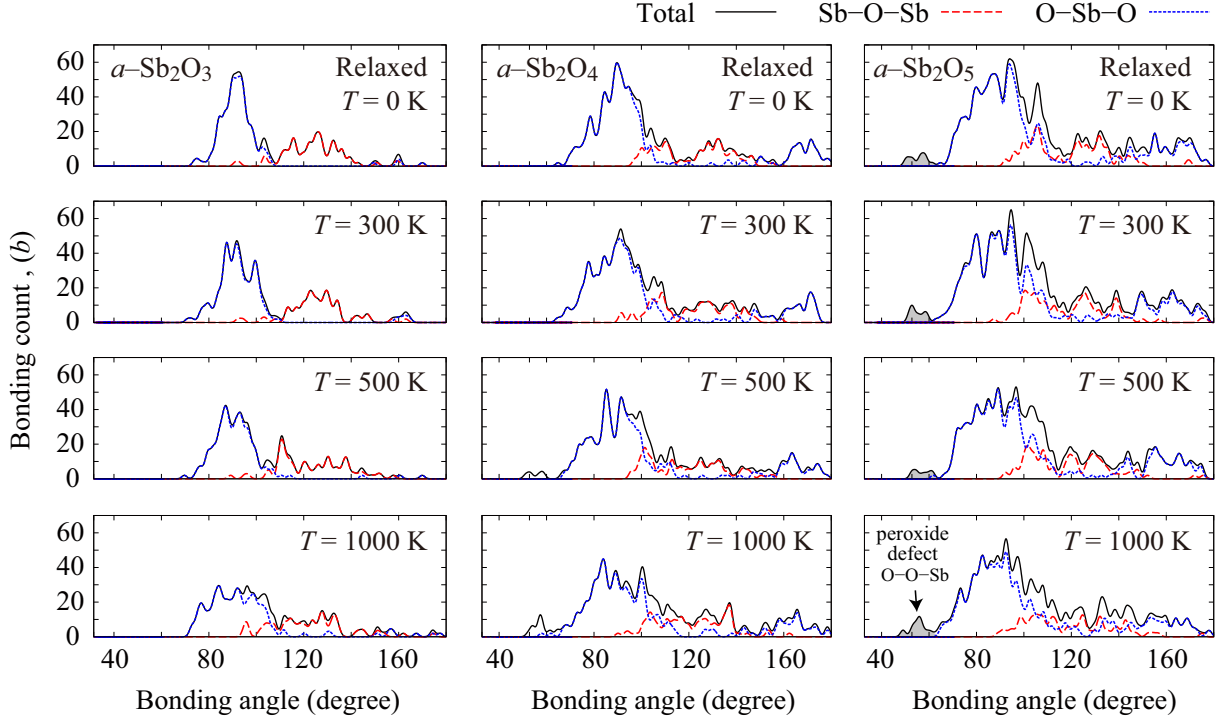}
\caption{(Color online) Bond-angle distributions of the amorphous antimony oxide models at different simulation temperatures. The finite-temperature data is obtained by characterizing snapshots from the MD trajectories, while the curves for the relaxed ($T$=0 K) models are computed from the quenched models after force minimization. The as-quenched sample {\it a\,}--Sb$_2$O$_3$ shows structural characteristics similar to valentinite (bond-angle distributions for the crystalline models are given in the supporting information), whereas after relaxation the distribution shifts closer to to 90$^{\circ}$.}
\label{figure03}
\end{figure}

The bond-angle distribution function provides further detail concerning the local geometry of the glass-network structure (Fig.\,\ref{figure03}; the distributions were computed for the same configurations as the (P)RDFs). Earlier work has suggested that the amorphous structure at low oxidation level (i.e. {\it a\,}--Sb$_2$O$_3$)is most similar to the valentinite polymorph ($\beta$ phase) of {\it c\,}--Sb$_2$O$_3$.\cite{terashima1996, som2010} In our amorphous Sb$_2$O$_3$ model, we also observed structural similarities with the $\beta$ phase, in particular a characteristic triple O--Sb--O peak in the bond-angle distribution function between 80$^{\circ}$ to 100$^{\circ}$ (see the T=300\,K panel for {\it a\,}--Sb$_2$O$_3$ in Fig.\,\ref{figure03}). 
This can be directly compared to the distribution of valentinite, which has characteristic peaks at 77.8$^{\circ}$, 91.2$^{\circ}$ and 97.1$^{\circ}$ (see Supporting Information). Note that experimental study of Johnson {\it et al.} found the O--Sb--O bonding angle of amorphous Sb$_2$O$_3$ sample is about 92$^{\circ}$, as in agreement to the bonding angle of the {\it a\,}--Sb$_2$O$_3$ in this study.

Upon relaxation, these peaks merge to a single broad feature, which is more in keeping with the $\alpha$ phase of {\it c\,}--Sb$_2$O$_3$. Our trioxide model thus contains structural feature from both the $\alpha$ and $\beta$ phases of {\it c\,}--Sb$_2$O$_3$, which may explain why Terashima {\it et al.},\cite{terashima1996} found that the vibrational spectra of the amorphous phase did not overlap exactly with those of either of the crystalline counterparts.

In {\it a\,}--Sb$_2$O$_4$, it is rather challenging to draw similarities between the amorphous structure and its crystalline counterparts, because the crystalline structures of $\alpha$ and $\beta$ polymorphs of {\it c\,}--Sb$_2$O$_4$ are structurally quite similar. One notable difference lies in the O--Sb--O bonding angle around the Sb(V) octahedral centers - $\alpha$--Sb$_2$O$_4$ displays a distorted geometry, leading to a peak at 167.5$^{\circ}$ between geometrically opposite O atoms, whereas the corresponding peak in the distribution for ($\beta$)--Sb$_2$O$_4$ lies at the ideal angle of 180$^{\circ}$. From the bond-angle distribution functions of our amorphous structures (see Fig.\,\ref{figure03}), once again we see a mixture of features from both crystalline phases, with continuous broad peaks ranging from 160$^{\circ}$ to 180$^{\circ}$ indicating the presence of both distorted and undistorted octahedral complexes.

The distribution functions of both {\it a\,}--Sb$_2$O$_4$ and {\it a\,}--Sb$_2$O$_5$ show evidence of distorted octahedral geometries, but there exists a notable difference between them. The $T = 0$\,K (i.e. relaxed) structure of {\it a\,}--Sb$_2$O$_4$ shows two broad yet clearly separated peaks in the high-angle part of the curve, indicating a mixture of motifs from both the $\alpha$ and $\beta$ phases. However, in the distribution calculated for the {\it a\,}--Sb$_2$O$_5$ model, there is a single continuous peak, indicating a range of distorted octahedral geometries with no particular bias towards certain bond angles.

The local geometrical arrangements, e.g. tetrahedral or octahedral geometries, could be more explicitly characterized if an appropriate analytical function was employed, e.g. as demonstrated by Debenedetti and co-workers.\cite{errington2001,chatterjee2008} However, the contracted tetrahedral bonding angle observed in the antimony oxides (92$^{\circ}$ from the work of Johnson {\it et al.}\cite{johnson2003b}, and 95.6$^{\circ}$ in our simulations) imposes a technical limitation in distinguishing tetrahedral and octahedral motifs in a manner similar to e.g. Refs.\,\onlinecite{errington2001} and \onlinecite{chatterjee2008}.

For the {\it a\,}--Sb$_2$O$_5$ system, the simulated melt-quench process led to the formation of a small number of peroxide defects, which is captured by characteristic weak short-range correlations in the PRDF curves in Fig.\,\ref{figure02}.
The peaks around 60$^{\circ}$ in the corresponding bond-angle distributions are also indicative of the formation of these defects (see Fig.\,\ref{figure03}). 
Peroxide species have been observed experimentally in other amorphous oxides, as is discussed in detail elsewhere.\cite{walsh2009}

The presence of peroxide defects marks a distinction between the amorphous phases with intermediate and high oxidation levels (i.e. {\it a\,}--Sb$_2$O$_4$, {\it a\,}--Sb$_2$O$_5$, respectively). In the former, there is some evidence for the presence of peroxide defects prior to quenching (see the $T=1000\,$K and $T=500\,$K curves for {\it a\,}--Sb$_2$O$_4$ in Figs.\,\ref{figure02} and \ref{figure03}). 
However, these defects are absent after quenching, and do not reform upon relaxation of the forces. 

The fact that these defects can form at high oxidation level suggests that Sb(V) can be reduced to Sb(IV/III) in the amorphous material, which in turn suggests that in a suitable form the material could catalyse redox processes (e.g. as nanoparticles or through surface sites). Crystalline Sb$_2$O$_3$ has been shown to be catalytically active, with the Sb(III) acting as a coordinating center during syntheses of polyethylene terephthalate.\cite{ravindranath1986} The role of Sb(III) during the catalysis is defined to be attracting oxygen of ester carbonyl groups, facilitating the oxygen to be more vulnerable to nucleophilic attack. However, it is not clear whether amorphous antimony oxides might also be catalytically active, e.g. due to surface Sb(III) centers comparable to {\it c\,}--Sb$_2$O$_3$ or via the possible Sb(V) $\leftrightarrow$ Sb(IV) interconversion, and a more detailed investigation is beyond the scope of the present study. However, this discussion is potentially interesting in light of the fact that amorphous catalysts are an active area of research in the community.\cite{morales-guio2014}

It is worth noting that in real systems peroxide defects may be oxidised and leave the material as molecular oxygen, for example during quenching or after vitrification, which is not possible in our periodic simulations. If this is the case, this would mean that amorphous compositions close to Sb$_2$O$_5$ may be difficult to prepare. Unfortunately, the limitations of the present simulations prohibit the study of additional compositions with non-integer stoichiometries, which would allow us to investigate this in more detail.

\begin{figure}[tb!]
\center
\includegraphics[width=0.60\textwidth]{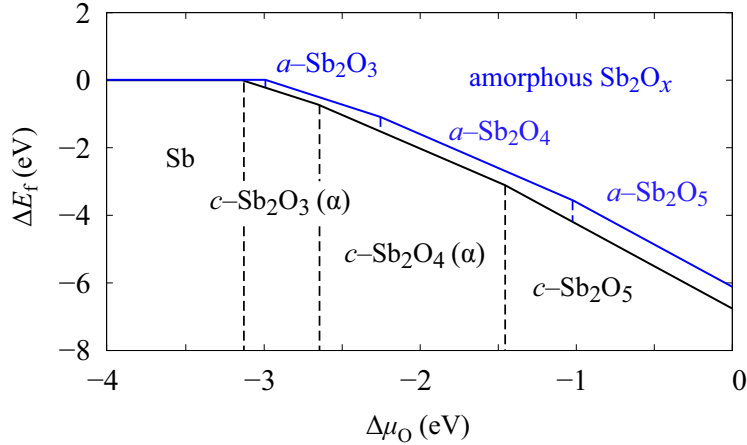}
\caption{(Color online) Calculated formation energies of amorphous Sb$_2$O$_x$, referenced to bulk Sb, with respect to the chemical potential of oxygen (referenced to the oxygen molecule). As expected, the amorphous phases are found to be metastable with respect to their crystalline counterparts. These results suggest that changing the stoichiometry of the amorphous phase would require a higher oxygen chemical potential doing so for the crystalline counterparts.}
\label{figure04}
\end{figure}

From the calculated total energies of the three relaxed amorphous models, we can construct a phase diagram for amorphous phase antimony oxide with respect to the chemical potential of oxygen (Fig.\,\ref{figure04}). 
The phase stability of glasses is frequently discussed from a kinetic point of view, but these results also confirm the thermodynamic metastability of the amorphous phases based on their formation energies. Compared to their crystalline counterparts, the amorphous phases show a significantly retarded oxidation-level change with respect to oxygen chemical potential (e.g. the oxidation level change from {\it a\,}--Sb$_2$O$_3$ to {\it a\,}--Sb$_2$O$_4$ requires a higher oxygen chemical potential than the transition between the {\it c\,}--Sb$_2$O$_3$ and {\it c\,}--Sb$_2$O$_4$). 

\begin{figure}[t!]
\center
\includegraphics[width=0.45\textwidth]{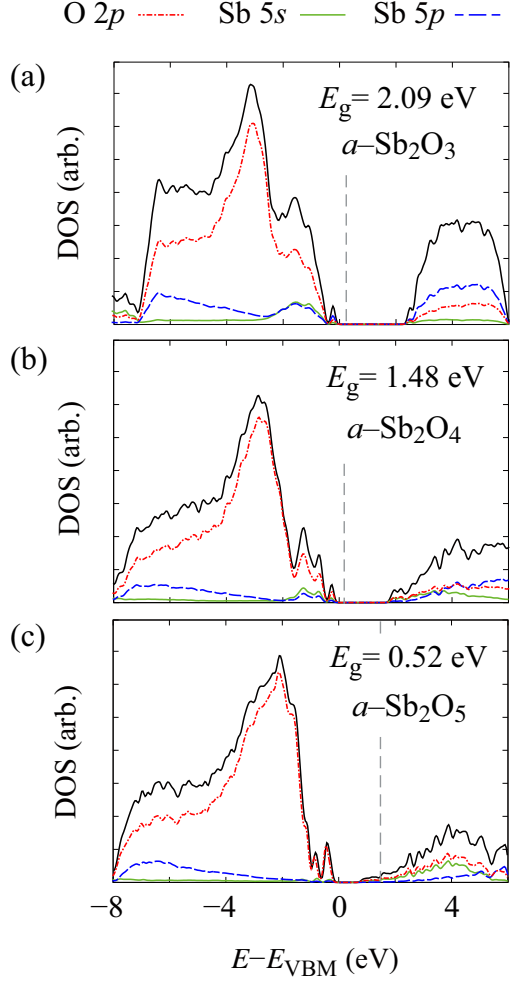}
\caption{(Color online) Calculated electronic density of states (DOS) curves for (a) {\it a\,}--Sb$_2$O$_3$, (b) {\it a\,}--Sb$_2$O$_4$, and (c) {\it a\,}--Sb$_2$O$_5$. The calculated Fermi level is shown as a vertical dashed line. For {\it a\,}--Sb$_2$O$_3$ and {\it a\,}--Sb$_2$O$_4$, a band gap is clearly evident, with the Fermi level lying closer to the valence-band maximum (VBM) than to the conduction-band minimum (CBM). In contrast, the {\it a\,}--Sb$_2$O$_5$ model shows strong $n$-type character, which appears to be due mainly to the presence of peroxide (O$_2^{2-}$) defects.}
\label{figure05}
\end{figure}

The electronic structure of amorphous antimony oxides is of particular importance for optoelectronic applications. 
Generally, the amorphous models in this study were found to have narrower energy gaps compared to their crystalline counterparts. We also observed a narrowing of the gap with increasing oxidation level (see Fig.\,\ref{figure05} and Tab.\,\ref{table01}), with the {\it a\,}--Sb$_2$O$_3$ system showing an energy gap of 2.09\,eV which decreases to 1.48\,eV and then to 0.52\,eV on going to {\it a\,}--Sb$_2$O$_4$ and then to {\it a\,}--Sb$_2$O$_5$. 

The origin of the narrowing energy gap is mainly due to the disordered amorphous structures. Amorphization of antimony oxides resulted in countable number of undercoordinated ions. This led to formation of localised states above the valence state edge. We confirmed this idea by matching the localised peaks of the projected DOS with the ions in the model.

Another reason of the narrowing energy gap is because the amorphous model has mixed phase comprising of the polymorphs of the crystalline counterparts. For the {\it a\,}--Sb$_2$O$_3$ as an example, the $\beta$ phase of its crystalline counterpart has a narrower energy gap compared to the $\alpha$ phase.

Comparing the calculated energy gap of the amorphous models to the available experimental data, we find the calculated energy gap of  {\it a\,}--Sb$_2$O$_3$ is smaller than the values reported in Wood {\it et al.}\cite{wood1972} and Terashima {\it et al.}\cite{terashima1996} One of the reason can be the use of the HSE06 functional was insufficient to fully recover the energy gap. However, this is less likely for the Sb$_2$O$_3$ because the use of HSE06 led to overestimation of the band gaps in this case. The measured value in Wood {\it et al,} for cubic antimony oxide, $c$--Sb$_2$O$_3$($\alpha$) phase, is about 4.0\,eV, while calculated value by HSE06 functional is 4.4\,eV. 

We discuss a few other possible origins of the difference between calculated energy gap and the measured values of the two experimental studies. Thermal energy at ambient condition (300\,K) or at an elevated temperature, might cause rearranging of the undercoordinated ions, leading to removal of the localised peaks above the valence state edge. Another cause can be due to recrystallization of the amorphous samples used in the experiments. For example, large number of discrete peaks begin to appear from nuclear magnetic resonance (NMR) results of the cited work,\cite{terashima1996} as the content of Sb$_2$O$_3$ exceeds 90\,\% of the B$_2$O$_3$--Sb$_2$O$_3$ glass matrix. 

The {\it a\,}--Sb$_2$O$_5$ DOS suggests $n$-type character, as indicated by the Fermi level lying close to the conduction band. This theoretical prediction is in line with experimental measurements,\cite{badawy1990} where $n$-type conductivity (2.5$\times$10$^{-4}\,$S/cm) of {\it a\,}--Sb$_2$O$_5$ thin films was reported. The DOS of {\it a\,}--Sb$_2$O$_5$ shows strongly localized states near the valance-band edge, mainly due to under-coordinated oxygen atoms in the glass matrix (see Fig.\,\ref{figure05}c). 

In general, amorphous antimony oxides are predicted to retain the large-gap semiconducting property as in their crystalline counterparts. The disordered structure resulted in narrowing of the energy gap compared to the crystallines. Compared to the $\alpha$, $\beta$, and $\gamma$ phases of {\it c\,}--Sb$_2$O$_3$, the gap of {\it a\,}--Sb$_2$O$_3$ is narrower by $2.3$, $1.07$, and $1.80$\,eV, respectively. The gap of {\it a\,}--Sb$_2$O$_4$ is narrower than those of the $\alpha$ and $\beta$ phases of {\it c\,}--Sb$_2$O$_4$ by $1.46$ and $1.65$\,eV, respectively, while the gap of {\it a\,}--Sb$_2$O$_5$ $1.28$\,eV smaller than that of {\it c\,}--Sb$_2$O$_5$. 

\begin{table}[tb!]
\centering
\caption{Calculated electronic energy gaps of amorphous and crystalline antimony oxides with different oxidation levels.}
\begin{tabular}{ccccc}
\noalign{\smallskip}\hline\noalign{\smallskip}
&Phase && Energy gap (eV)& \\\noalign{\smallskip}\hline\noalign{\smallskip}\hline\noalign{\smallskip}
&{\it a\,}--Sb$_2$O$_3$ && 2.09 & \\\noalign{\smallskip}
&{\it a\,}--Sb$_2$O$_4$ && 1.48 & \\\noalign{\smallskip}
&{\it a\,}--Sb$_2$O$_5$ && 0.52 & \\\noalign{\smallskip}\noalign{\smallskip}
&{\it c\,}--Sb$_2$O$_3$($\alpha$) && 4.40 & \\\noalign{\smallskip}
&{\it c\,}--Sb$_2$O$_3$($\beta$) && 3.17 & \\\noalign{\smallskip}
&{\it c\,}--Sb$_2$O$_3$($\gamma$) && 3.61 & \\\noalign{\smallskip}
&{\it c\,}--Sb$_2$O$_4$($\alpha$) && 2.94 & \\\noalign{\smallskip}
&{\it c\,}--Sb$_2$O$_4$($\beta$) && 3.13 & \\\noalign{\smallskip}
&{\it c\,}--Sb$_2$O$_5$ && 1.80 \\\noalign{\smallskip}\hline\noalign{\smallskip}
\end{tabular}
\label{table01}
\end{table}

\section{Conclusion}
In summary, we have employed simulations based on density-functional theory to explore the material structure and properties of single-component antimony-oxide glasses, {\it a\,}--Sb$_2$O$_x$, at oxidation levels of $x = 3, 4, 5$. Radial-distribution functions and bond-angle distribution functions have been used to characterize the local structures, and suggest that the glass networks contain a mixture of the local geometries found in the crystalline counterparts, with some evidence for the formation of peroxide defects at high oxidation levels. The thermodynamic metastability of the amorphous phases compared to their crystalline counterparts is evident from calculated phase diagrams. Our calculations were found to be in good agreement with existing experimental results. Finally, the calculated HSE06 electronic density-of-states yielded fundamental energy gaps smaller than those of the crystalline counterparts. These results also suggest the $n$-type conductivity of $a$--Sb$_2$O$_5$ as indicated by the shift of Fermi level. The theoretical results presented here should serve as an important reference point for future studies of the amorphous phases of single-component antimony oxides, which belong to the technologically-relevant class of heavy metallic-oxide glasses.

\begin{acknowledgments}
We gratefully acknowledge support from the Basic Science Research Program by the NRF (Grant No. 2014R1A1A1003415). JMS is supported by a UK Engineering and Physical Sciences Research Council Programme Grant (grant no. EP/K004956/1). Computational resources have been provided by the KISTI supercomputing center (KSC-2015-C3-009) and the Archer HPC facility, which was accessed through the UK Materials Chemistry Consortium, funded by EPSRC grant no. EP/L000202.
\end{acknowledgments}

%

\end{document}